\documentclass[a4paper,oneside]{article}
\usepackage[T1]{fontenc}
\usepackage{syntonly}
\usepackage[ansinew]{inputenc}

\usepackage{tocbibind}
\usepackage{amsmath}
\usepackage{amssymb}
\usepackage{physics}
\usepackage{multirow}
\usepackage{subcaption}
\usepackage{threeparttable}
\usepackage{longtable}
\usepackage{animate}
\usepackage{cite}
\usepackage{authblk}
\usepackage{morefloats}

\usepackage{subcaption}
\usepackage{caption}
\DeclareCaptionLabelSeparator{bar}{~\rule[-0.4ex]{0.2ex}{1em}~}
\DeclareCaptionLabelFormat{subfor}{\textbf{#2}}
\newcommand*\bfcaption[2]{\caption[#1]{\textbf{#1.}#2}}
\usepackage{bibentry}
\usepackage[pdftex]{color,graphicx}
\usepackage[update,prepend]{epstopdf}

\author[1,2]{Ricard Alert\thanks{ricard.alert@princeton.edu}}
\affil[1]{Princeton Center for Theoretical Science, Princeton University, Princeton, USA, NJ 08544}
\affil[2]{Lewis-Sigler Institute for Integrative Genomics, Princeton University, Princeton, USA, NJ 08544}
\author[3,4,5,6]{Xavier Trepat\thanks{xtrepat@ibecbarcelona.eu}}
\affil[3]{Institute for Bioengineering of Catalonia, The Barcelona Institute for Science and Technology (BIST), Barcelona, Spain, 08028}
\affil[4]{Facultat de Medicina, University of Barcelona, Barcelona, Spain, 08036}
\affil[5]{Instituci\'{o} Catalana de Recerca i Estudis Avan\c{c}ats (ICREA), Barcelona, Spain, 08028}
\affil[6]{Centro de Investigaci\'{o}n Biom\'{e}dica en Red en Bioingenier\'{i}a, Biomateriales y Nanomedicina, Barcelona, Spain, 08028}
\title{Physical Models of Collective Cell Migration}
\date{\today}
\usepackage{xcolor}
\definecolor{UBcolor}{HTML}{007CC1}
\usepackage[colorlinks=true,pdfnewwindow=true,linkcolor=UBcolor,citecolor=UBcolor,urlcolor=UBcolor,breaklinks=true,linktocpage]{hyperref}
\usepackage[all]{hypcap}
\usepackage[capitalise,nameinlink]{cleveref}
\begin{document}

\maketitle

\begin{abstract}
Collective cell migration is a key driver of embryonic development, wound healing, and some types of cancer invasion. Here we provide a physical perspective of the mechanisms underlying collective cell migration. We begin with a catalogue of the cell-cell and cell-substrate interactions that govern cell migration, which we classify into positional and orientational interactions. We then review the physical models that have been developed to explain how these interactions give rise to collective cellular movement. These models span the sub-cellular to the supracellular scales, and they include lattice models, phase fields models, active network models, particle models, and continuum models. For each type of model, we discuss its formulation, its limitations, and the main emergent phenomena that it has successfully explained. These phenomena include flocking and fluid-solid transitions, as well as wetting, fingering, and mechanical waves in spreading epithelial monolayers. We close by outlining remaining challenges and future directions in the physics of collective cell migration.
\end{abstract}

\clearpage

\tableofcontents

\section{Introduction} \label{intro}

It has long been recognized that cells move as collectives during development, regeneration and wound healing. Reports from the late XIXth century already agreed that these processes involve collective movements of cells but mechanisms remained controversial\cite{Born1897,Holmes1914,Herrick1932,Vaughan1966}. Some authors proposed that cell movements were driven by pressure, either pre-existing in the tissue or generated de-novo by cell division (see Ref. \cite{Herrick1932} and references therein). Others claimed that cells would move by spreading their volume to occupy the largest possible surface\cite{Born1897}. Yet, others defended that cell sheets advanced by the active pulling force generated by leader cells at the tissue margin\cite{Holmes1914}. In those early days of cell biology, proposed mechanisms were largely physical in origin but not even the sign of the tissue stress, i.e. tension vs compression, was agreed upon. Later on, the discovery of genes and proteins shifted the attention from the whole to the parts, and the search for a global physical understanding of collective migration was largely abandoned.

\begin{figure*}[tb]
\begin{center}
\includegraphics[width=0.9\textwidth]{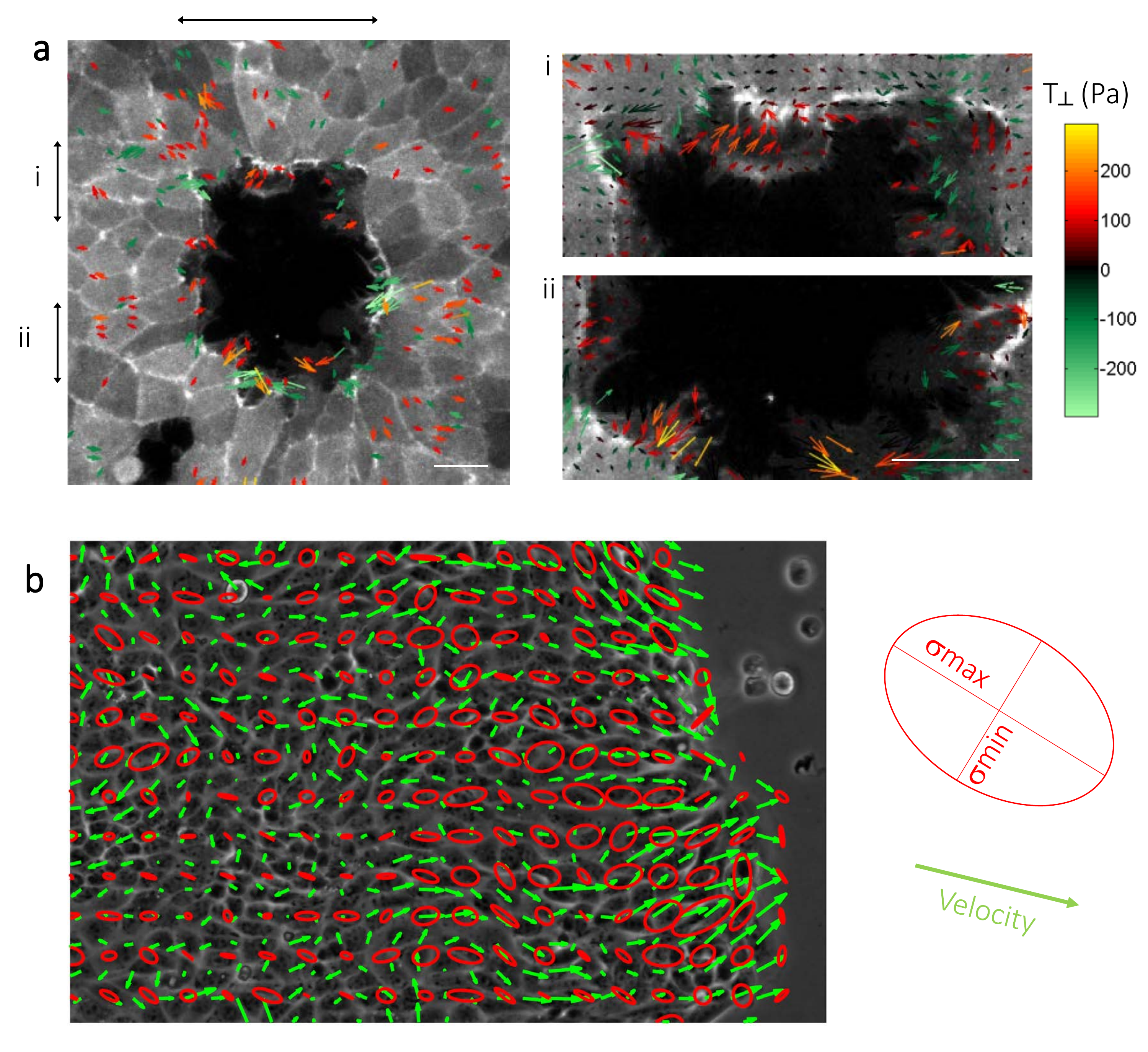}
\end{center}
  {\phantomsubcaption\label{Fig measurements-a}}
  {\phantomsubcaption\label{Fig measurements-b}}
\bfcaption{Mechanical measurements during collective migration of cell monolayers}{ \subref*{Fig measurements-a}, Vectorial representation of traction forces in LifeAct-GFP MDCK cells closing a wound. Color coding indicates the value of the radial component, with positive forces pointing away from the wound. For clarity, values between 100 and -100 Pa were not plotted. Panels labeled as i and ii show a close-up of the regions indicated by arrows in panel \subref*{Fig measurements-a}. Scale bar, 20 $\mu$m. Adapted from \cite{Brugues2014}. \subref*{Fig measurements-b}, Velocity vectors (green) and monolayer stress ellipses (red) indicating the maximum and minimum principal stresses in an expanding colony of MDCK cells (phase contrast). Adapted from \cite{Serra-Picamal2012}.} \label{Fig old}
\end{figure*}

This trend has been reversed in the past decade due to groundbreaking technical\cite{Roca-Cusachs2017} and conceptual\cite{Marchetti2013,Prost2015,Julicher2018} advances together with a progressive questioning of reductionist approaches\cite{Good2018}. Time-lapse imaging and fluorescence microscopy have become standard tools in life-science laboratories, and technologies such as particle imaging velocimetry (borrowed from fluid mechanics) now enable a detailed mapping of velocity fields and strain tensors in the tissue\cite{Poujade2007,Serra-Picamal2012}. Moreover, a range of new technologies such as traction microscopy have enabled the direct mapping of the forces that cells exert on their surroundings as they migrate\cite{duRoure2005,Trepat2009}. All mechanical variables relevant to the problem of collective cell migration have thus become available in time and space (\cref{Fig old}). This technological revolution has coincided with the development of the theory of active matter\cite{Marchetti2013,Prost2015,Julicher2018}, which provides an ideal framework to rationalize the collective movement of cells. The traditional view that physics should serve to illuminate biological function is shifting towards the idea that biological systems inspire new physical theories and allow us to test them; the concept of `physics for biology' is now paralleled by the emergent notion of `biology for physics'. In this sense, from the perspective of condensed matter physics, collective cell migration is interesting as a prominent example of the emergence of collective mechanical phenomena in a system of soft active entities with complex interactions. Finally, life scientists have recognized that collective cell migration is not only key to development, regeneration and wound healing, but also to devastating diseases such cancer\cite{Friedl1995}. The coincidence in time of these different technological and conceptual advances has placed collective cell migration back to the center stage of research at the interface between life and physical sciences.

Collective cell migration comes in different flavors depending on the biological tissue and process\cite{Friedl2009}. During epithelial morphogenesis, wound healing and regeneration, cells generally move as sheets adhered on an inert hydrogel called the extracellular matrix (ECM). In some forms of cancer invasion, cells also invade as sheets at the interface between tissues. In general, however, both in development and in tumor invasion, cells invade as strands or clusters within a complex 3D environment composed of ECM and different cell types\cite{Friedl2009,Cai2014,Clark2015}. Despite recent advances\cite{Campas2016}, we remain far from accessing physical forces in 3D, so we will focus this review on cell sheets migrating in 2D (\cref{Fig old}). In this mode of migration, central to epithelial function, all relevant cellular forces have been accurately mapped in vitro and can therefore be used to test physical models. For other perspectives on this subject, the reader is referred to excellent recent reviews\cite{Hakim2017,Ladoux2017,Xi2019}.

\section{Forces and interactions of migrating cells} \label{forces}

In this section, we propose a classification of the forces and interactions that a migrating cell exerts on and experiences from other cells and the substrate. In addition to distinguishing cell-cell and cell-substrate interactions, we separate the interactions that directly act on a cell's position from those that affect cell orientation.
Even though any categorization may suffer from some degree of oversimplification, we think that it nevertheless provides some unifying principles over a large body of somehow fragmented literature.


\subsection{Positional cell-substrate interactions} \label{positional-CS}

The field of continuum mechanics defines a traction as a force per unit area applied at any point of the surface of a body. In cell mechanics, traction is usually understood as the stress applied by a cell on its underlying inert substrate, typically a polymeric gel known as extracellular matrix (ECM). Cell-substrate traction can be interpreted as the sum of two contributions: the force that drives cell motion, which we call active traction, and the passive friction between the cell and its substrate.

\subsubsection{Active traction forces} \label{CS-active}

Active traction forces stem from the cell's actomyosin cytoskeleton, where the action of myosin molecular motors on actin filaments generates contractile forces. These forces are then transmitted to the substrate through cell-substrate adhesion complexes called focal adhesions, which physically connect the cytoskeleton to the extracellular matrix (\cref{Fig interactions}). For traction forces to lead to cell motion, the cell must break symmetry and polarize to define a front and a rear. To do so, the cell often develops frontal actin-based protrusions such as lamellipodia and filopodia. These structures generate an inwards-pointing active traction that is most prominent at the cell's leading edge (\cref{Fig interactions}). The resulting force propels the cell forward in the direction of its polarity, defined by the position of the protrusion with respect to the cell's center of mass.\footnote{The front-rear polarity of a cell is a morphological, dynamical, and biochemical asymmetry between the cell's leading and trailing edges. The polarity direction of an isolated cell can be unambiguously identified from the direction of its migration, regardless of its sub-cellular origin. However, this may no longer be true in a tissue, where cell motion is also affected by intercellular forces. In this case, to distinguish polarity from velocity, one must identify a sub-cellular observable that defines cell polarity. Morphological features such as protrusions are often not apparent in cells inside tissues, which may feature cryptic lamellipodia that extend beneath neighboring cells. Thus, an alternative approach is to identify cell polarity from traction forces. This is valid as long as tractions are dominated by active forces, with negligible contributions from passive friction forces. Identifying sub-cellular features that appropriately account for cell polarity remains an open challenge.} Thus, in models, the force that drives cell migration is usually assumed proportional to a cell polarity variable, with a coefficient that depends both on cell-substrate adhesion and on the active force-generating processes in the cytoskeleton.

\subsubsection{Cell-substrate friction forces} \label{CS-friction}

\begin{figure*}[tb]
\begin{center}
\includegraphics[width=\textwidth]{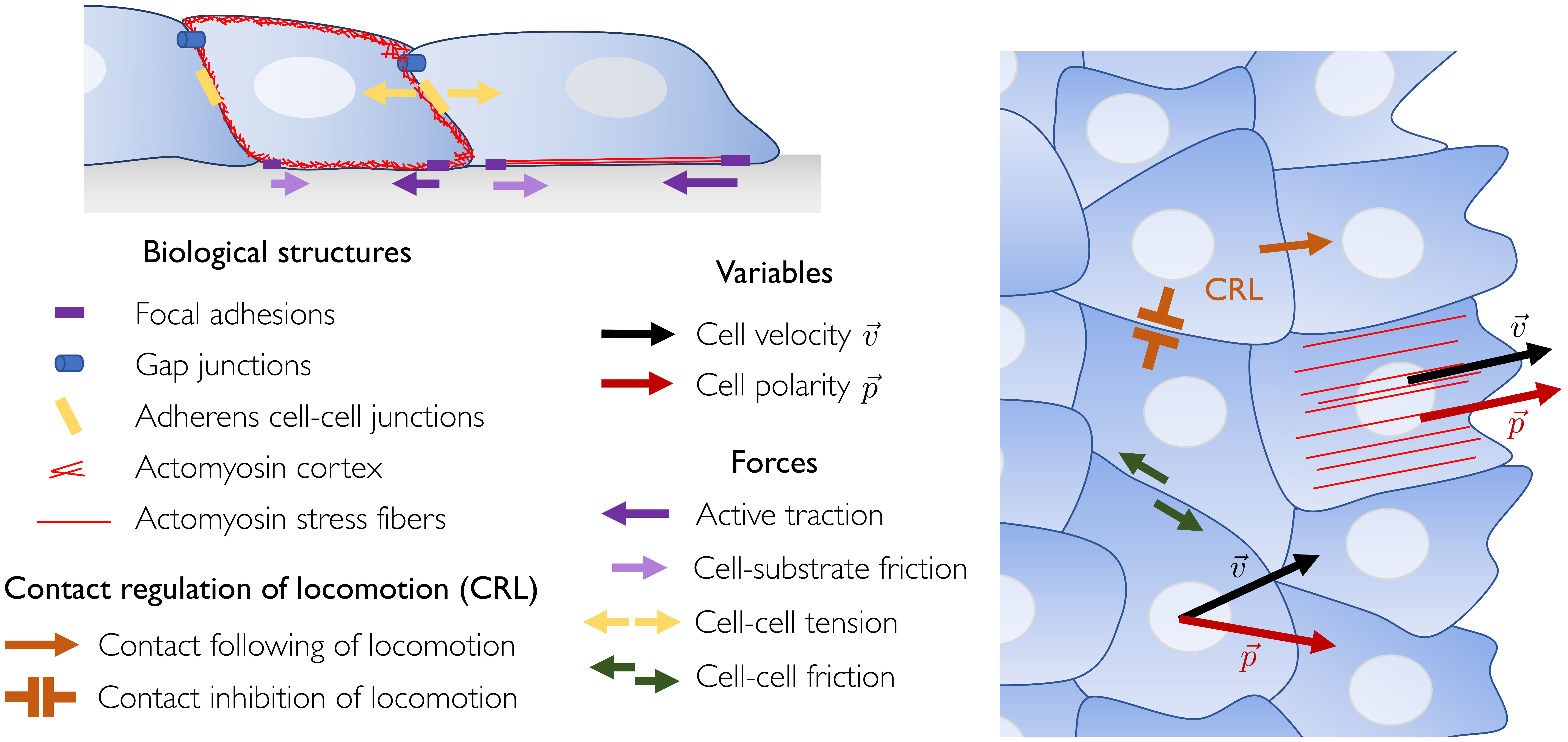}
\end{center}
\bfcaption{Forces and interactions of migrating cells}{ Schematic representation of a migrating epithelial monolayer. Side view (left), top view (right). We illustrate a subset of the interactions discussed in \cref{forces}. In particular, we sketch some of the biological structures that generate and transmit cell-substrate and cell-cell forces. We also indicate contact interactions that regulate and coordinate cell migration, as well as physical variables used to describe collective cell migration.} \label{Fig interactions}
\end{figure*}

Cell motion takes place at very low Reynolds numbers, which implies that inertial forces are negligible and, hence, that the resultant force on the cell's center of mass must vanish. Indeed, the active traction applied by the cell on the substrate is balanced by friction forces. Friction with the surrounding fluid medium is usually negligible in front of cell-substrate friction forces, which are mediated by the attachment and detachment of proteins at focal adhesions (\cref{Fig interactions}). On time-scales larger than the inverse binding rates, this protein-mediated friction is expected to be proportional to the velocity of the cell relative to the substrate \cite{Schwarz2013}. Thus, in a first approximation, cell-substrate friction is often modeled as a viscous damping force akin to Stokes' drag, with a coefficient that reflects cell-substrate adhesion. However, the kinetics of focal adhesion proteins under force are extremely non-linear and involve reinforcement feedbacks that can be accounted for in more detailed models of cellular friction \cite{Elosegui-Artola2018b}.

\subsection{Positional cell-cell interactions} \label{positional-CC}

\subsubsection{Cell-cell adhesion} \label{CC-adhesion}

A characteristic feature of epithelial cells is that they establish stable cell-cell adhesions, whereas mesenchymal cells tend to form transient and weaker adhesions. Cell-cell adhesion is mediated by specific transmembrane protein complexes, which build cell-cell junctions that physically link the actomyosin cortices of the adhering cells, thereby enabling force transmission between cells (\cref{Fig interactions}). Cell-cell junctions endow tissues with cohesion energy and surface tension, as well as with a bulk modulus that quantifies their resistance to rapid isotropic expansions, which is mainly due to cytoskeletal elasticity \cite{Khalilgharibi2016,Xi2019}. Thus, different modeling frameworks account for cell-cell adhesion by either an interfacial energy contribution, a short-range attraction that opposes cell-cell detachment, or a tissue bulk modulus. In addition, by enabling the transmission of active cytoskeletal tension, cell-cell adhesion is also an implicit factor in tissue-scale active stress terms.

The influence of cell-cell adhesion on tissue mechanical properties does not end here. On the one hand, decreasing cell-cell adhesion leads to less elongated cell shapes that can induce a jamming transition whereby the tissue acquires a finite shear modulus, thus becoming a solid material \cite{Bi2015,Park2015a}. Therefore, cell-cell adhesion may not only provide bulk but also shear elasticity to epithelial tissues. On the other hand, cell-cell adhesion proteins turn over and, hence, cell-cell junctions are remodeled. Junction remodeling is a source of dissipation that can relax stress, possibly contributing to the long-time viscous response of fluid tissues \cite{Khalilgharibi2016,Xi2019,Blanch-Mercader2017,Iyer2019}.

\subsubsection{Cell-cell friction} \label{CC-friction}

Cell-cell adhesion also entails cell-cell friction forces when cells slide past each other (\cref{Fig interactions}). Similar to cell-substrate friction, cell-cell friction is based on the sliding, turnover, and attachment kinetics of cell-cell junction proteins \cite{Czirok2013,Peglion2014,Garcia2015}. Usually, cell-cell friction is modeled as a shear force proportional to the relative velocity between the cells or, in tissue-level descriptions, as shear viscous stresses.

\subsubsection{Cell-cell repulsion} \label{compressibility}

In addition to the attractive interactions mediated by cell-cell adhesion, adhered cells also experience a soft repulsion from other cells. At short times, cell compression is resisted by cytoskeletal elasticity \cite{Khalilgharibi2016,Xi2019}, which, in epithelial monolayers, gives rise to an area compressibility. At longer times, however, both the cytoskeleton and cell-substrate adhesions may reorganize to enable cell shape changes. If their volume is conserved, cells under compression can lose area and gain height, at least until their nuclei become tightly packed. However, cells can actually change their volume by exchanging fluid both with the surrounding medium through water channels and with other cells through cell-cell channels called gap junctions \cite{Zehnder2015,Zehnder2015a} (\cref{Fig interactions}). Finally, under sufficient compression, epithelial cells can be extruded from a monolayer \cite{Kocgozlu2016,Saw2017,Chen2018}, thus enabling monolayer area reduction. The opposite process, cell insertion in a monolayer, also occurs in spreading cell aggregates \cite{Beaune2014}. Altogether, this means that epithelial monolayers can change their area via dissipative processes, with an associated viscosity.

\subsubsection{Active cell-cell forces} \label{CC-active}

A last type of cell-cell forces are active forces generated by myosin molecular motors in the cytoskeleton and transmitted through cell-cell junctions (\cref{Fig interactions}). Cytoskeletal structures such as the cell cortex and the apical actin belt generate a roughly isotropic tension at the cell scale, thus giving rise to isotropic active stress at the tissue level. However, migrating cells are polarized and, hence, their cytoskeleton exhibits highly anisotropic structures such as stress fibers (\cref{Fig interactions}). Stress fibers generate anisotropic tension that, in addition to being transmitted to the substrate as traction forces (\cref{CS-active}), can also be transmitted to cell-cell junctions through the cell cortex, thus giving rise to anisotropic active stresses at the tissue scale. Given that they are similarly generated, traction forces and anisotropic cell-cell active stresses are interdependent \cite{Maruthamuthu2011}. Yet, they are distinct since their respective transmission to the ECM and neighboring cells relies on different adhesion complexes (\cref{Fig interactions}).

\subsection{Orientational cell-cell interactions} \label{orientational-CC}

\subsubsection{Polarity alignment} \label{CC-alignment}

One of the most prominent orientational interactions between cells is the tendency to align their polarities. Alignment might simply rely on the elongated and deformable shape of migrating cells, but it may also involve biochemical regulation of cell migration. Polarity alignment is often explicitly implemented either via Vicsek-like rules or via torques on cell polarity in discrete models, and via an orientational stiffness of the polarity field in continuum models.

\subsubsection{Contact regulation of locomotion} \label{CRL}

Here, we propose the term \textit{contact regulation of locomotion} (CRL) to subsume a number of processes whereby cells tune their migration direction upon contact interactions with other cells (\cref{Fig interactions}). At least three such processes have been described. First, contact inhibition of locomotion (CIL) refers to the process whereby, upon head-to-head collision, many cell types retract their lamellipodia and repolarize in a different direction, thus migrating away from cell-cell contacts \cite{Stramer2017,Mayor2016}. Second, contact following of locomotion (CFL) refers to the tendency of cells to follow others upon head-to-tail contact \cite{Li2018,Fujimori2019}. Finally, in addition to altering the migration direction, head-to-tail collisions have also been found to increase the persistence of cell motion --- a tendency known as contact enhancement of locomotion (CEL) \cite{dAlessandro2017}. In general, CRL depends strongly on the cell-cell collision angle \cite{Desai2013,George2017}, thus making orientational cell-cell interactions highly anisotropic. The mechanisms underlying CRL may be diverse and, given that cells polarize in response to tension transmitted through cell-cell junctions \cite{Desai2009,Weber2012,Ng2012,Roca-Cusachs2013,Das2015,Ladoux2016,Vishwakarma2018}, they could rely on mechanotransduction of cell-cell forces \cite{Davis2015,Scarpa2015,Roycroft2016,Roycroft2018}. Although a number of phenomenological models of CRL have been proposed, a coherent theoretical picture of contact regulation of locomotion is lacking.

\subsubsection{Polarity-flow alignment} \label{polarity-flow}

Inhomogeneous tissue flows may produce shear. Similar to molecules in liquid crystals, elongated cells subject to shear should experience a torque that tends to minimize shear stress. Indeed, shear tissue flows reorient cell polarity in the fly wing \cite{Aigouy2010} as well as the cell division axis in epithelial monolayers \cite{Marel2014}. Moreover, cells in epithelial monolayers tend to migrate in the local direction of lowest shear stress --- a behavior known as plithotaxis \cite{Tambe2011,Trepat2011,He2015,Zaritsky2015}. However, unlike in ordinary liquid crystals, cell reorientation may not entirely stem from cell shape but it likely involves an active mechanosensitive response. Regardless of its yet-unclear mechanism, this feedback between polarity and flow mediates another type of orientational cell-cell interactions. Even though some recent continuum models have probed the effects of flow-polarity coupling \cite{Blanch-Mercader2017c,Duclos2018}, further research is needed to clarify its role in collective cell migration.


\subsubsection{Polarity-shape alignment} \label{polarity-shape}

Almost by definition, cell polarity and cell shape asymmetry are interdependent and, hence, they are often assumed to align. Then, given that cell-cell interactions modify cell shape, cell-autonomous polarity-shape alignment can give rise to intercellular alignment interactions \cite{Coburn2013,Lober2015,Barton2017}.

\subsection{Orientational cell-substrate interactions} \label{orientational-CS}

\subsubsection{Polarity-velocity alignment} \label{polarity-velocity}

Through their interaction with the substrate, cells may be able to align their polarity to their velocity, thus tending to align self-propulsion with drag cell-substrate forces \cite{Peyret2018}. Such a polarity-velocity coupling is a generic property of active polar systems interacting with a substrate \cite{Brotto2013,Kumar2014a,Oriola2017,Maitra2019}. For these systems, the polarity not only reorients in flow gradients but also in uniform flows, like a weathercock in the wind. Even though its cellular mechanism is not yet well understood, polarity-velocity alignment has been introduced in a number of models. However, in some situations, polarity and velocity are strongly misaligned in epithelial monolayers \cite{Kim2013,Notbohm2016} (\cref{Fig interactions}), possibly due to the dominance of cell-cell interactions \cite{Notbohm2016,Zimmermann2016}. How such possibly conflicting polarization cues coexist and cooperate remains poorly understood \cite{Lin2018}.

\subsubsection{Substrate-induced polarization} \label{substrate-polarity}

In addition to polarizing in response to cell-cell forces (\cref{CRL}), cells can also polarize in response to asymmetric forces at the cell-substrate interface \cite{Bun2014,Ladoux2016}. In particular, given that cells exert larger tractions on more adhesive and/or stiffer substrates, gradients of substrate adhesivity and/or stiffness can polarize cells \cite{Breckenridge2014,Marzban2018,Saez2007}. The ensuing migrations towards regions of higher adhesivity and/or stiffness are known as haptotaxis and durotaxis, respectively. Moreover, even changes in uniform substrate properties may lead to cell polarization. Specifically, increasing substrate stiffness triggers an isotropic-nematic transition in the actomyosin cytoskeleton \cite{Gupta2015,Gupta2016,Ladoux2016,Gupta2019a}. This transition results in cell elongation, which, in turn, might promote a spontaneous polarization \cite{Prager-Khoutorsky2011}.





\section{Physical models, from sub-cellular to supracellular scales} \label{models}

In this section, we review the different physical descriptions that have been used to model collective cell migration. These descriptions cover different levels of coarse-graining; we start from those describing sub-cellular detail and move up to continuum models that only describe supracellular features. Complementary presentations have been provided in recent reviews \cite{Camley2017a,Hakim2017,Spatarelu2019}. Here, we emphasize how the cellular forces and interactions reviewed in the previous section can be accounted for by each of the modeling approaches. We focus on two-dimensional models that explicitly include cell migration. 


\subsection{Lattice models: The cellular Potts model} \label{lattice}

In the spirit of classical models of statistical mechanics, such as the paradigmatic Ising model, lattice models describe individual cells as domains on a lattice, thus resolving sub-cellular details of cell shape (\cref{Fig CPM-a}). In particular, this description is based on the Potts model, and hence it is known as the Cellular Potts Model (CPM) \cite{Graner1992}. Each lattice site $i=1,\ldots,N$ is assigned a state variable $\sigma_i=1,\ldots,m$ corresponding to one of $m-1$ cells. The state of each lattice site is then updated using a state-exchange Monte Carlo scheme with Metropolis dynamics at a sufficiently low temperature to ensure that cells remain as compact domains.

\subsubsection{Effective Hamiltonian} \label{effective-Hamiltonian-CPM}

\begin{figure*}[tbp]
\begin{center}
\includegraphics[width=\textwidth]{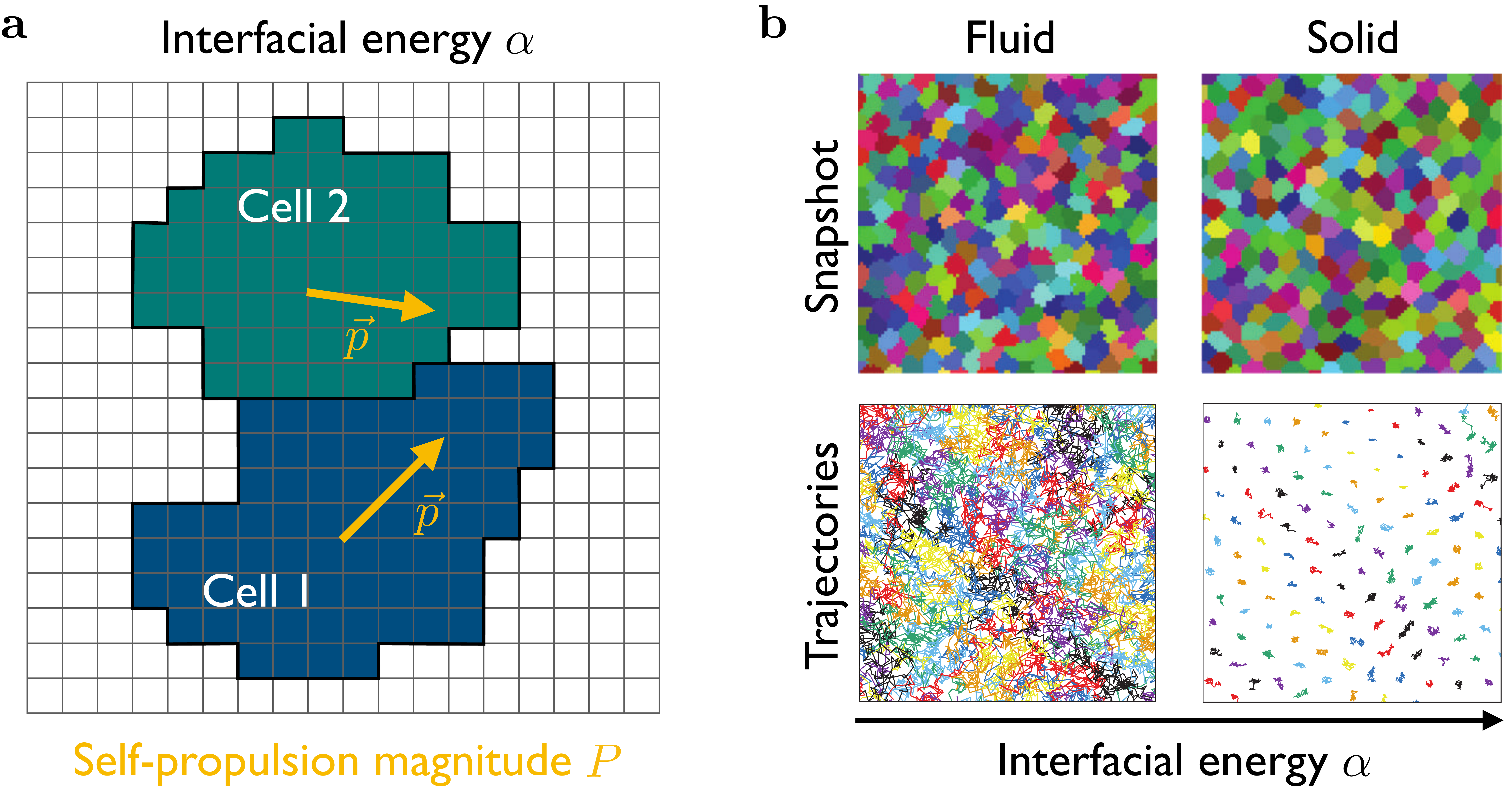}
\end{center}
  {\phantomsubcaption\label{Fig CPM-a}}
  {\phantomsubcaption\label{Fig CPM-b}}
\bfcaption{Lattice Models: The Cellular Potts Model}{ \subref*{Fig CPM-a}, Lattice sites corresponding to two different cells are shown in different colors. Cell-cell and cell-medium interfaces have an interfacial energy $\alpha$ (black). Cell migration in the direction of the polarity is favored by a self-propulsion magnitude $P$. \subref*{Fig CPM-b}, Snapshots of the system and cell trajectories at the fluid ($\alpha=1$) and solid ($\alpha=4$) regimes. Neighboring cells are colored differently, with arbitrary colors. Cell boundaries are rougher and longer for smaller interfacial energy. Adapted from \cite{Chiang2016} with permission from EPL.} \label{Fig CPM}
\end{figure*}

The dynamics minimizes the effective Hamiltonian
\begin{equation} \label{eq CPM-Hamiltonian}
\mathcal{H} = \sum_{\langle i,j\rangle} J(\sigma_i,\sigma_j) + \lambda \sum_{\sigma=1}^{m-1} (A_\sigma - A_0)^2 - P \sum_{\sigma=1}^{m-1} \vec{R}_\sigma\cdot \vec{p}_\sigma.
\end{equation}
The original CPM only included the first two terms. The first term, whose sum runs over neighboring sites $\langle i,j\rangle$, accounts for the interfacial tension between neighboring cells as well as between cells and the medium (state $\sigma=m$), which are encoded in the interaction matrix $J$. The simplest choice is $J(\sigma_i,\sigma_j)= \alpha (1-\delta_{\sigma_i,\sigma_j})$, where $\alpha$ is the interfacial energy that controls the amplitude of cell shape fluctuations (\cref{Fig CPM}). Thus, this energy captures the combined effects of cell-cell adhesion and cortical tension (\cref{CC-adhesion,CC-active}). The second term penalizes changes in cell area around a preferred value $A_0$, with an area modulus $\lambda>0$ (\cref{Fig CPM-a,compressibility}). The area of cell $\sigma$, i.e. its number of lattice sites, is simply given by $A_\sigma = \sum_{i=1}^N \delta_{\sigma_i,\sigma}$.

\subsubsection{Cell migration} \label{cell-migration-CPM}

Later, the third term was added to implement cell motility by decreasing the energy of those configurations in which a cell's center of mass $\vec{R}_\sigma = A_\sigma^{-1}\sum_{i=1}^N (x_i,y_i) \, \delta_{\sigma_i,\sigma}$ has advanced toward the direction of its polarity $\vec{p}_\sigma$ \cite{Rappel1999}. This term corresponds to an active polar force of magnitude $P>0$ on each cell: $\vec{F}_\sigma = -\vec{\nabla}_{\vec{R}_\sigma} \mathcal{H} = P \vec{p}_\sigma$ (\cref{Fig CPM-a,CS-active}). The model does not specifically include friction forces. Rather, an effective damping of cell motion arises from the Metropolis dynamics itself, which is dissipative in nature. In fact, the mean cell speed is linear in $P/\alpha$ over a wide range of parameter values \cite{Kabla2012}. Thus, the cell-cell coupling strength $\alpha$, which controls the diffusion coefficient of a cell in the absence of motility, is proportional to the effective viscous friction coefficient, consistent with the Stokes-Einstein relation.

\subsubsection{Polarity dynamics} \label{polarity-dynamics-CPM}

The cellular polarity $\vec{p}_\sigma$ was proposed to align with the velocity over some time-scale \cite{Szabo2010,Kabla2012} (\cref{polarity-velocity}) or, alternatively, to simply undergo rotational diffusion \cite{Chiang2016}. In a variant of the CPM, cell motion was dictated by the gradient of a self-secreted chemoattractant, whose concentration evolves with its own dynamics \cite{Ouaknin2009}. CPMs with alternative polarity dynamics should be explored in the future. Along these lines, Coburn et al. have recently proposed a hybrid CPM that accounts for CIL \cite{Coburn2018} (\cref{CRL}).

\subsubsection{Collective phenomena} \label{collective-CPM}

Initially, the self-propelled CPM was primarily used to study velocity correlations of complex flows in cell monolayers \cite{Szabo2010,Kabla2012,Czirok2013}. More recently, it has also been used to study fluid-solid transitions and glassy dynamics in cell monolayers \cite{Kabla2012,Chiang2016} (\cref{Fig CPM-b}), collective rotations \cite{Segerer2015}, gap closure \cite{Coburn2018}, and tissue spreading \cite{Thueroff2019}, including the fingering instability of the tissue front \cite{Ouaknin2009,Thueroff2019}.

\subsubsection{Discussion} \label{discussion-CPM}

The CPM is based on an explicit and detailed description of cell shape and cell-cell adhesion which, by means of intensive simulations, enables close investigation of cell-scale mechanisms of cell rearrangements. However, the Metropolis dynamics yields somewhat artificial cell shape fluctuations that depend on a temperature parameter not directly related to experimental measurements. Moreover, the model is not readily suited to incorporate some kinds of cellular interactions relevant for collective cell migration. In particular, how to distinguish cell-cell and cell-substrate friction and how to appropriately capture the active nature of some cellular forces with the relaxational algorithm of the CPM remains unclear.


\subsection{Phase-field models} \label{phase-field}

\begin{figure*}[tb]
\begin{center}
\includegraphics[width=\textwidth]{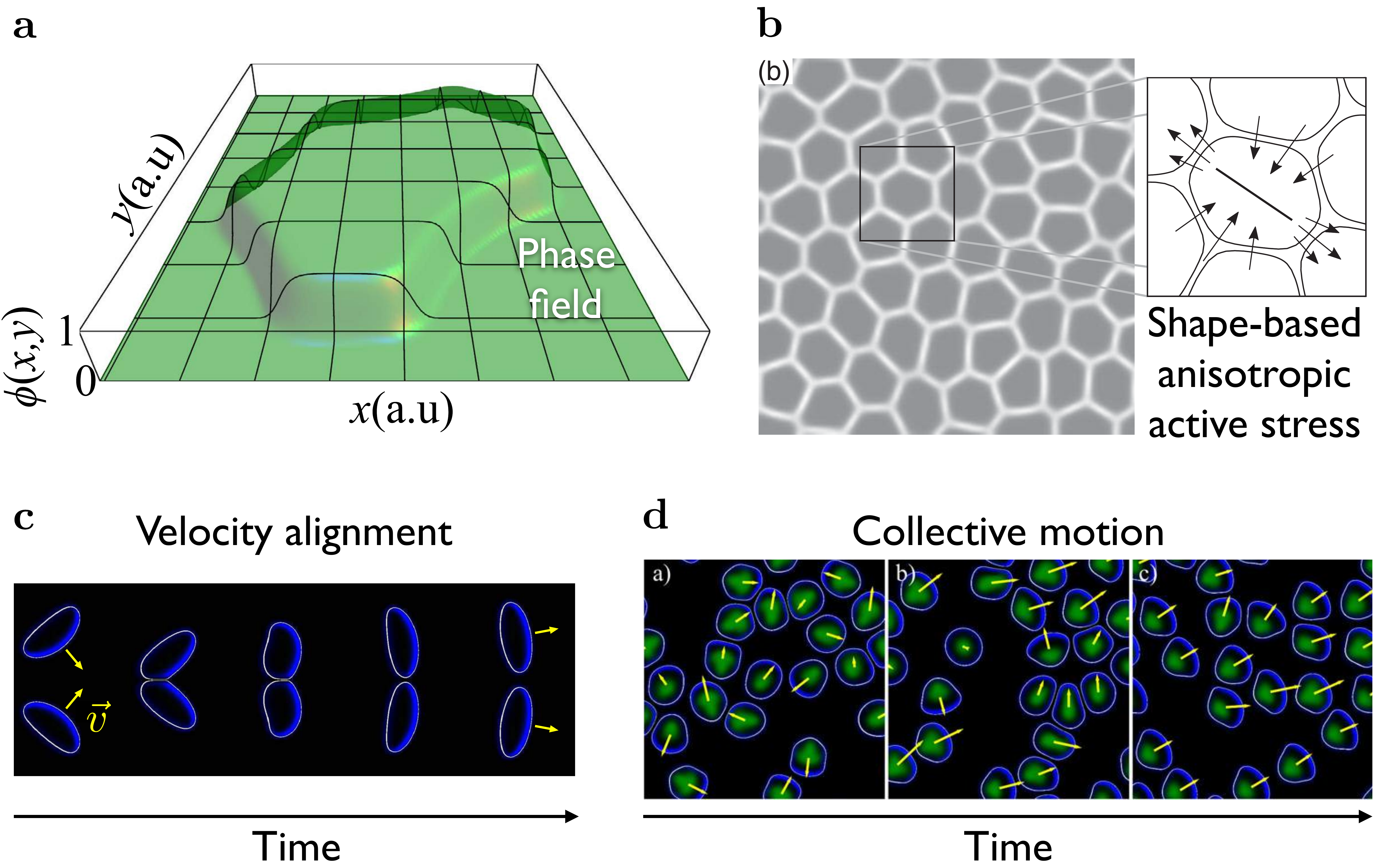}
\end{center}
  {\phantomsubcaption\label{Fig phase-field-a}}
  {\phantomsubcaption\label{Fig phase-field-b}}
  {\phantomsubcaption\label{Fig phase-field-c}}
  {\phantomsubcaption\label{Fig phase-field-d}}
\bfcaption{Phase-Field Models}{ \subref*{Fig phase-field-a}, Phase field of a cell. Adapted from \cite{Palmieri2015}. \subref*{Fig phase-field-b}, The overlap between phase fields, $\sum_{i\neq j} \phi_i\phi_j$, identifies cell-cell interfaces (white). The inset shows cell contours, $\phi_i=1/2$, along with a sketch of extensile stress along the principal axis of the cell deformation tensor (see text). Adapted with permission from \cite{Mueller2019}. Copyright (2019) by the American Physical Society. \subref*{Fig phase-field-c}-\subref*{Fig phase-field-d}, Collisions between deformable cells lead to velocity alignment (\subref*{Fig phase-field-c}) and collective motion (\subref*{Fig phase-field-d}). Adapted from \cite{Lober2015}.} \label{Fig phase-field}
\end{figure*}

With their origins in interface dynamics \cite{Gonzalez-Cinca2004}, phase-field models also describe cell shape in sub-cellular detail. However, unlike the CPM, they do not rely on a lattice. Rather, each cell $i=1,\ldots,N$ is described by a phase field $\phi_i(\vec{r},t)$, which is $1$ inside the cell and $0$ outside (\cref{Fig phase-field-a}). A similar approach relies on describing cell shape via a contour function \cite{Coburn2013}. Some models describe even intracellular structures, such as the nucleus, using additional phase fields \cite{Camley2014a}.

\subsubsection{Phase-field free energy} \label{phase-field-free-energy}

Cell-cell interactions are built into a free energy functional of the phase field. Although formulations vary \cite{Camley2014a,Palmieri2015,Mueller2019}, a possible form is $\mathcal{F}= \mathcal{F}_{\textrm{CH}} + \mathcal{F}_{\textrm{area}} + \mathcal{F}_{\textrm{cell-cell}}$ with
\begin{subequations}
\begin{align}
\mathcal{F}_{\text{CH}} &= \sum_{i=1}^N \frac{\gamma}{\epsilon} \int_{\mathcal{A}} \left[ 4\phi_i^2 (1-\phi_i)^2 + \epsilon^2 |\vec{\nabla}\phi_i|^2 \right] \dd^2\vec{r},\\
\mathcal{F}_{\text{area}} &= \sum_{i=1}^N \mu \left(1-\frac{1}{\pi R^2} \int_{\mathcal{A}} \phi_i^2 \,\dd^2\vec{r}\right)^2,\\
\mathcal{F}_{\text{cell-cell}} &= \sum_{i=1}^N \sum_{j\neq i} \frac{\kappa}{\epsilon} \int_{\mathcal{A}} \left[\phi_i^2 \phi_j^2 - \tau\epsilon^4 |\vec{\nabla}\phi_i|^2 |\vec{\nabla}\phi_j|^2\right] \dd^2\vec{r}.
\end{align}
\end{subequations}
The first contribution is a Cahn-Hilliard free energy that stabilizes the phase-field interface. The first term is a double-well potential with minima at the cell interior ($\phi_i=1$) and exterior ($\phi_i=0$), which are connected by an interface of width $\epsilon$ and tension $\gamma$ that delineates the cell boundary. Here, we have neglected the bending rigidity of the interface \cite{Camley2014a}. The second contribution penalizes departures of cell area from its preferred value $\pi R^2$, with area modulus $\mu$ (\cref{compressibility}). The third contribution accounts for cell-cell interactions. It includes a repulsive term that penalizes cell overlapping (\cref{compressibility}), with strength $\kappa$, and an attractive interaction between cell boundaries that models cell-cell adhesion (\cref{CC-adhesion}), with strength $\kappa\tau$ \cite{Camley2014a}.

\subsubsection{Phase-field dynamics and force balance} \label{phase-field-dynamics}

The dynamics of cell shape reads
\begin{equation}
\partial_t \phi_i + \vec{v}_i \cdot\vec{\nabla}\phi_i = -\frac{\delta \mathcal{F}}{\delta \phi_i}.
\end{equation}
Here, $\vec{v}_i$ is a cell velocity defined as \cite{Palmieri2015}
\begin{equation}
\vec{v}_i = \frac{1}{\xi} \int_{\mathcal{A}} \frac{\delta \mathcal{F}}{\delta \phi_i}\vec{\nabla}\phi_i\,\dd^2\vec{r} = \frac{1}{\xi} \int_{\mathcal{A}} \frac{\delta \mathcal{F}_{\textrm{cell-cell}}}{\delta \phi_i}\vec{\nabla}\phi_i\,\dd^2\vec{r} = \frac{1}{\xi} \vec{F}_i^{\textrm{int}},
\end{equation}
where $\xi$ is a friction coefficient (\cref{CS-friction}), and $\vec{F}_i^{\textrm{int}}$ is an interaction force on the interface of cell $i$ due to overlaps with neighboring cells. This relationship can be generalized to include cell motility in the form of an active polar contribution $T_a \vec{p}_i$ to the force balance \cite{Peyret2018} (\cref{CS-active}):
\begin{equation} \label{eq force-balance-phase-field}
\xi \vec{v}_i = \vec{F}^{\textrm{int}}_i + T_a \vec{p}_i,
\end{equation}
The interaction force $F_i^{\textrm{int}}$ can also be generalized to account for additional interactions. In continuum mechanics, short-range interaction forces are described in terms of the stress tensor field $\mathbf{\sigma}(\vec{r},t)$. In the phase-field formulation, this corresponds to \cite{Mueller2019}
\begin{equation} \label{eq interfacial-force-phase-field}
\vec{F}_i^{\textrm{int}} = \int_{\mathcal{A}} \phi_i \, \vec{\nabla}\cdot \mathbf{\sigma}\,\dd^2\vec{r} = - \int_{\mathcal{A}} \mathbf{\sigma} \cdot \vec{\nabla}\phi_i \,\dd^2\vec{r}.
\end{equation}
In addition to the the phase-field interactions, which give a pressure term,
the stress tensor may also include other contributions such as viscous and active stresses. In this case, combining the approaches of Refs. \cite{Mueller2019,Peyret2018}, the stress tensor could read
\begin{equation} \label{eq stress-tensor-phase-field}
\mathbf{\sigma}(\vec{r},t) = -P(\vec{r},t)\,\mathbb{I} + \xi_c \sum_{j\neq i} (\vec{v}_i-\vec{v}_j) \, \vec{\nabla} \phi_j(\vec{r},t) - \zeta \mathbf{Q}(\vec{r},t),
\end{equation}
where the second term accounts for cell-cell friction with coefficient $\xi_c$ (\cref{CC-friction}), and the third term describes anisotropic active stresses proportional to the nematic order parameter tensor field $\mathbf{Q}(\vec{r},t) = \sum_{i=1}^N \phi_i(\vec{r},t)\, \mathbf{S}_i$ (\cref{CC-active}). Here, $\mathbf{S}_i$ is the orientation tensor of cell $i$, which may be based either on cells' polarities, $\mathbf{S}_i = \vec{p}_i \vec{p}_i - 1/2\,|\vec{p}_i|^2\, \mathbb{I}$, or on cells' shapes as proposed in \cite{Mueller2019}, $\mathbf{S}_i = -  \int_{\mathcal{A}} \left[ (\vec{\nabla}\phi_i)^T\, \vec{\nabla}\phi_i - |\vec{\nabla}\phi_i|^2 \,\mathbb{I}\right]\dd^2\vec{r}$ (\cref{Fig phase-field-b}).

\subsubsection{Polarity dynamics} \label{polarity-dynamics-phase-field}

Regarding the polarity dynamics, interactions such as CIL and CFL (\cref{CRL}), polarity alignment (\cref{CC-alignment}), and polarity-velocity alignment (\cref{polarity-velocity}) have been explored \cite{Camley2014a}, as well as couplings to chemotactic fields \cite{Najem2016}. More recently, an alignment of cell polarity toward the direction of the total interfacial force has also been implemented \cite{Peyret2018} (\cref{CRL}).

\subsubsection{Collective phenomena} \label{collective-phenomena-phase-field}

Phase-field models have primarily addressed the emergence of collective motion from cell-cell interactions. Whereas some works focused on explicit orientational interactions \cite{Camley2014a}, other studies showed that, when cell polarity is coupled to cell shape asymmetry (\cref{polarity-shape}), collisions between deformable cells lead to cell-cell velocity alignment and collective motion \cite{Coburn2013,Lober2015} (\cref{Fig phase-field-c,Fig phase-field-d}). Recently, the phase-field model has been employed to explain the emergence of extensile nematic behavior \cite{Mueller2019} and to recapitulate collective velocity oscillations \cite{Peyret2018} in epithelial monolayers.

\subsubsection{Discussion} \label{discussion-phase-field}

The phase-field formalism provides a detailed description of cell shape while tackling some of the issues of the CPM. Foremost, it introduces a force balance \cref{eq force-balance-phase-field} that provides a physical dynamics, thus going beyond the energy minimization process of the CPM, which imposes static mechanical equilibrium at each step. Moreover, the phase-field model is currently better connected to tissue mechanics (\cref{eq interfacial-force-phase-field}), and it can explicitly account for cell-cell and cell-substrate friction as well as for active stresses (\crefrange{eq force-balance-phase-field}{eq stress-tensor-phase-field}) \cite{Mueller2019,Peyret2018}.

\subsection{Active network models} \label{network}

With precedents in the physics of foams \cite{Weaire1999}, network models describe epithelial tissues as networks of polygonal cells \cite{Alt2017}. Thus, albeit in less detail than lattice and phase-field models, these models still describe sub-cellular features of cell shape. They encompass two subtypes of models: vertex and Voronoi models.

\subsubsection{Vertex and Voronoi models} \label{vertex-Voronoi}

\begin{figure*}[tb]
\begin{center}
\includegraphics[width=\textwidth]{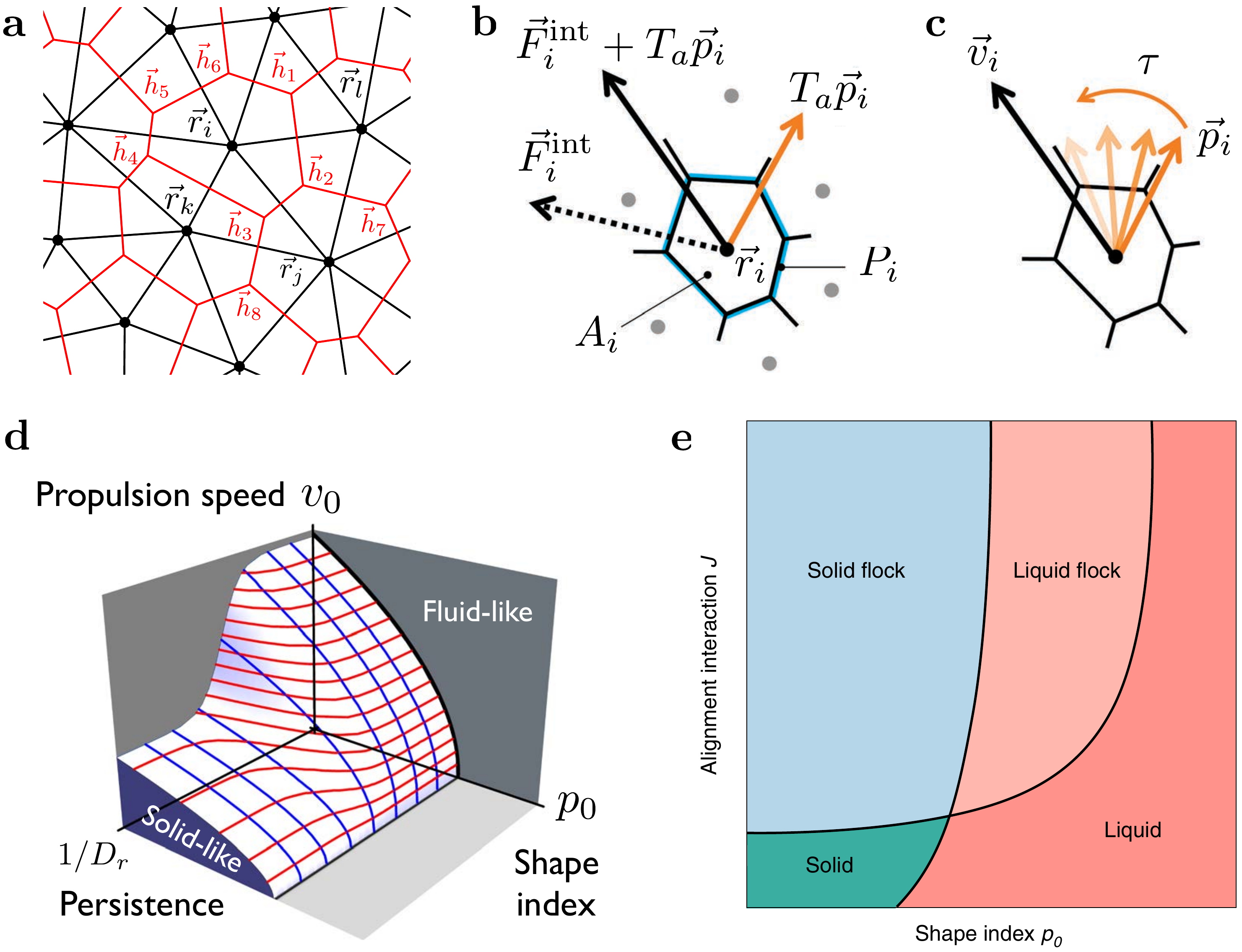}
\end{center}
  {\phantomsubcaption\label{Fig SPV-a}}
  {\phantomsubcaption\label{Fig SPV-b}}
  {\phantomsubcaption\label{Fig SPV-c}}
  {\phantomsubcaption\label{Fig SPV-d}}
  {\phantomsubcaption\label{Fig SPV-e}}
\bfcaption{Active Network Models}{ \subref*{Fig SPV-a}, Cell centers at positions $\{\vec{r}\}$ are connected by the Delaunay triangulation (black). Its dual is the Voronoi tessellation (red) that defines cell boundaries and vertices at positions $\{\vec{h}\}$. Reprinted from \cite{Bi2016}. \subref*{Fig SPV-b}, Cells are parametrized by an area $A_i$ and a perimeter $P_i$, and experience a self-propulsion force $T_a \vec{p}_i$ (orange) and an interaction force $\vec{F}_i^{\textrm{int}}=-\vec{\nabla}_{\vec{r}_i}\mathcal{F}$ (dashed black), which give the resultant force (black). Adapted from \cite{Giavazzi2018} with permission from The Royal Society of Chemistry. \subref*{Fig SPV-c}, Polarity-velocity alignment with a time-scale $\tau$. Adapted from \cite{Giavazzi2018} with permission from The Royal Society of Chemistry. \subref*{Fig SPV-d}, Schematic phase diagram of the fluid-solid transition in the SPV model in terms of the shape index $p_0=P_0/\sqrt{A_0}$, and the self-propulsion speed $v_0=T_a/\xi$ and persistence $D_r^{-1}$. Adapted from \cite{Bi2016}. \subref*{Fig SPV-e}, Schematic phase diagram of the SPV model with polarity-velocity alignment at rate $J=\tau^{-1}$ (see \subref*{Fig SPV-c}). Reprinted from \cite{Trepat2018a} by permission from Springer Nature.} \label{Fig SPV}
\end{figure*}

In vertex models, the degrees of freedom are the vertices of the polygons. Alternatively, the network can be described by the cell centers, which reduces the number of degrees of freedom. These descriptions are known as Voronoi models because, given the positions of the cell centers, the cell-cell boundaries are delineated by the Voronoi tessellation (\cref{Fig SPV-a}). The difference in the number of degrees of freedom has important consequences for the mechanical properties of the network, which may thus differ between vertex and Voronoi models \cite{Sussman2018a}. Moreover, cell motion, as well as cell division and cell death, may entail topological rearrangements of the network of cell-cell interfaces. In Voronoi models, the network is dynamic, evolving with each recomputation of the tessellation. In vertex models, in contrast, network rearrangements entail the appearance and disappearance of vertices, which requires implementing specific rules.

\subsubsection{Energy function} \label{energy-network}

In both descriptions, as in the previous approaches, cellular properties and interactions are encoded in an energy function, usually parametrized in terms of the areas $A_a$ and perimeters $P_a$ of cells $a=1,\ldots,N$:
\begin{equation} \label{eq vertex-energy}
\mathcal{F} = \sum_{a=1}^N \left[\frac{\kappa}{2} (A_a - A_0)^2 + \Lambda P_a + \frac{\Gamma}{2} P_a^2\right].
\end{equation}
Here, $\kappa$ is the modulus of cell area around its preferred value $A_0$ (\cref{compressibility}). Respectively, $\Lambda = \gamma_c - w/2$ is the line tension of the cell-cell interfaces that connect the vertices, which results from the coaction of the cortical tension along cell-cell contacts, $\gamma_c$, and the cell-cell adhesion energy $w$ (\cref{CC-adhesion,CC-active}) \cite{Manning2010,Winklbauer2015}. When cell-cell adhesion dominates, the line tension $\Lambda$ becomes negative and the cell-cell interface tends to expand. This expansion is eventually saturated by other cellular processes. This saturation is encoded in the third term of \cref{eq vertex-energy}, which gives rise to a perimeter-dependent line tension. This term is a key difference between models of tissues and foams; for the latter, $\Lambda$ is always positive and the quadratic perimeter term is absent \cite{Weaire1999}. The two perimeter contributions in \cref{eq vertex-energy} can be recasted as an energetic penalty for departures from a preferred perimeter $P_0 = -\Lambda/\Gamma$:
\begin{equation}
\mathcal{F} = \sum_{a=1}^N \left[\frac{\kappa}{2} (A_a - A_0)^2 + \frac{\Gamma}{2} (P_a - P_0)^2\right].
\end{equation}

\subsubsection{Cell migration and force balance} \label{force-balance-network}

Cell motility can then be implemented by applying active polar forces either on the vertices or on the cell centers, giving rise to Active Vertex Models (AVM) \cite{Salm2012,Lin2018,Mathur2018,Schaumann2018,Staddon2018} and Self-Propelled Voronoi models (SPV) \cite{Li2014,Bi2016}, respectively. Thus, the corresponding degrees of freedom $i = 1,\ldots, n$ move according to \cref{eq force-balance-phase-field}, albeit with $\vec{F}_i^{\textrm{int}} = -\vec{\nabla}_{\vec{r}_i} \mathcal{F}$ (\cref{Fig SPV-b}). In addition, to account for interfacial effects at tissue boundaries, Salm and Pismen added a `wetting force' at the tissue edge \cite{Salm2012}, whereas Barton et al. included surface tension and bending forces \cite{Barton2017}.


\subsubsection{Polarity dynamics} \label{polarity-dynamics-network}

The most popular orientational interaction in SPV models has been polarity-velocity alignment \cite{Li2014,Malinverno2017,Barton2017,Giavazzi2018,Petrolli2019} (\cref{Fig SPV-c,polarity-velocity}). However, polarity-shape alignment \cite{Barton2017,Mathur2018} (\cref{polarity-shape}), polarity alignment \cite{Barton2017} (\cref{CC-alignment}), CIL \cite{Lin2018,Mathur2018} and force-induced polarization \cite{Salm2012} (\cref{CRL}), as well as couplings to self-secreted chemoattractants \cite{Salm2012} have also been considered. Coburn et al. have proposed a hybrid model that accounts for CIL and polarity-shape alignment \cite{Coburn2016}.

\subsubsection{Collective phenomena} \label{collective-phenomena-network}

Using the SPV model, Bi et al. studied how cell motility modifies the solid-fluid transition displayed by passive vertex models, showing that both self-propulsion speed and persistence favor the fluid phase \cite{Bi2016} (\cref{Fig SPV-d}). Other studies have focused on the onset of collective motion, showing that cell-autonomous polarity-velocity alignment (\cref{Fig SPV-c,polarity-velocity}) gives rise to emergent cell-cell alignment, which leads to coherent rotations \cite{Li2014} and flocking \cite{Barton2017,Malinverno2017,Giavazzi2018}. Altogether, SPV models predict four distinct phases: solid, liquid, solid flock, and liquid flock (\cref{Fig SPV-e}). The solid phase supports elastic collective oscillations excited by self-propulsion \cite{Barton2017,Petrolli2019}.



\subsubsection{Discussion} \label{discussion-network}

By construction, most network models describe confluent tissues, in which cells are packed without free space between them. Therefore, these models are restricted to collective migration of epithelial cell groups. This limitation has been addressed in recent work that generalizes the Voronoi model to non-confluent tissues \cite{Teomy2018}. In general, network models are particularly suited to study the role of cell geometry and topological rearrangements on cell motion. Similar to the CPM, a current limitation of network models is that they account for neither internal dissipation nor anisotropic active stresses in the tissue. Recent efforts to include cell-cell friction \cite{Koride2018} and to relate network geometry to the tissue stress tensor \cite{Yang2017b} offer possible ways to address these limitations.


\subsection{Particle models} \label{particle}

Particle models are rooted in the physics of particulate media such as granular materials. Compared to previous descriptions, particle models resolve even less details of cell shape by treating each cell as one or two circular particles. Using two particles still allows capturing cell shape anisotropy \cite{Basan2013}, and even a head-tail asymmetry if the particles have different size \cite{Schnyder2017}. Otherwise, details of cell shape are entirely overlooked.

\subsubsection{Cell-cell interaction potential} \label{potential-particle}

\begin{figure*}[tb]
\begin{center}
\includegraphics[width=\textwidth]{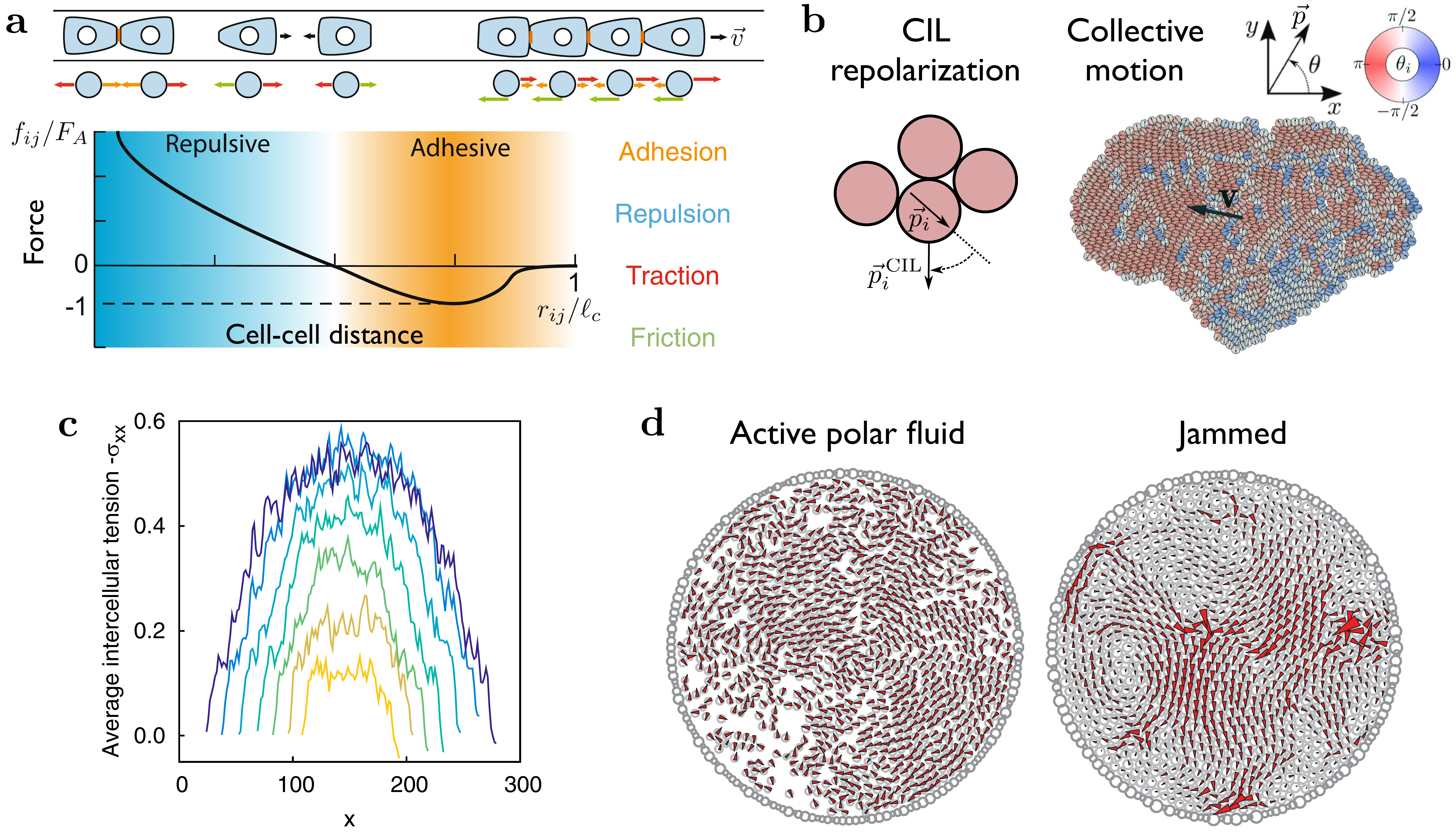}
\end{center}
  {\phantomsubcaption\label{Fig particles-a}}
  {\phantomsubcaption\label{Fig particles-b}}
  {\phantomsubcaption\label{Fig particles-c}}
  {\phantomsubcaption\label{Fig particles-d}}
\bfcaption{Particle Models}{ \subref*{Fig particles-a}, Schematic representation of forces in particle models. In different situations, cells experience various amounts of cell-cell repulsion (blue) and adhesion (orange), and cell-substrate active traction (red) and passive friction (green). Adapted from \cite{George2017}. \subref*{Fig particles-b}, In a simple model, CIL rotates cell polarity $\vec{p}_i$ towards the direction $\vec{p}_i^{\,\textrm{CIL}}$ pointing away from cell-cell contacts. With these interactions, a cell cluster can spontaneously polarize and undergo collective motion, as indicated by the center-of-mass velocity $\vec{v}$. The color code indicates the cell polarity angle $\theta_i$. Adapted from \cite{Smeets2016}. \subref*{Fig particles-c}, In growing tissues, CIL gives rise to tension profiles similar to experimental measurements. Reprinted from \cite{Zimmermann2016}. \subref*{Fig particles-d}, Increasing the packing fraction of confined self-propelled particles leads to jamming. The jammed phase supports undamped low-frequency modes. Red arrows indicate cell velocity. Adapted with permission from \cite{Henkes2011}. Copyright (2011) by the American Physical Society.} \label{Fig particles}
\end{figure*}

Positional cell-cell interactions are implemented via a central interparticle potential $V(|\vec{r}_i - \vec{r}_j|)$. As for the other kinds of energy functions, no general principle predicts the exact form of the potential. Rather, simple forms are often proposed on phenomenological grounds. Typically, the potential features a short-range repulsion, which usually includes a hard core to prevent cell overlaps (\cref{compressibility}). However, to capture cell extrusion, a recent model proposed a soft-core repulsion with a finite energy plateau \cite{Smeets2016}. In addition to the repulsive part, the potential often features a mid-range attraction to account for cell-cell adhesion (\cref{Fig particles-a,CC-adhesion}). Unless modeling biochemical signaling or substrate-mediated elastic interactions, long-range non-contact interactions are not included and, hence, the potential is cut off at the maximal cell radius.

\subsubsection{Cell migration and force balance} \label{force-balance-particle}

As in other approaches, cell motility is often accounted for by an active polar force $T_a \vec{p}_i$ on the particles. In addition to central and self-propulsion forces, cell-substrate viscous friction $-\xi\vec{v}_i$ and cell-cell friction with coefficient $\xi_c$ can also be added to the force balance (\cref{Fig particles-a}). Thus, the equation of motion of cell $i=1,\ldots,N$ reads
\begin{equation} \label{eq force-balance-particles}
\xi \vec{v}_i = T_a \vec{p}_i + \sum_{\langle i,j\rangle} \left[ - \vec{\nabla}_{\vec{r}_i} V(|\vec{r}_i - \vec{r}_j|) + \xi_c [\vec{v}_i - \vec{v}_j] \right].
\end{equation}
Here, the sum is restricted to contacting cells. In addition, to account for interfacial phenomena, Tarle et al. added surface tension along the tissue edge as well as a coupling of the motility force to the edge curvature \cite{Tarle2015}.

\subsubsection{Polarity dynamics} \label{polarity-dynamics-particle}

A wide range of polarity interactions have been studied in particle models. Interestingly, as in phase-field and SPV models, the combination of short-range forces with autonomous polarity-velocity alignment (\cref{polarity-velocity}) can lead to cell-cell velocity alignment and flocking \cite{Szabo2006,Basan2013}.
Nevertheless, other studies have explicitly implemented Vicsek-like velocity alignment rules \cite{Sepulveda2013}. Other polarity interactions that have been modeled include CIL \cite{Bindschadler2007,Desai2013,Woods2014,Zimmermann2016,Smeets2016,Schnyder2017,George2017,Copenhagen2018}, CFL \cite{Desai2013}, and force-induced repolarization \cite{George2017} (\cref{CRL}).

\subsubsection{Collective phenomena} \label{collective-phenomena-particle}

Particle models have unveiled that not only polarity-velocity alignment but also CRL interactions can give rise to collective motion \cite{Desai2013,Woods2014,Smeets2016,George2017}. In particular, because it tends to anti-align the polarities of cell pairs, CIL would not be expected to lead to a state with net polarity. However, in cell clusters, CIL induces a coupling between the polarity and density fields that gives rise to a spontaneous symmetry breaking towards collective motion \cite{Smeets2016} (\cref{Fig particles-b}).
Moreover, adjusting CIL strength to the local concentration of a chemoattractant enables intrinsically collective modes of chemotaxis \cite{Camley2016}. Particle models have also shown that CIL can stabilize cell monolayers against dewetting \cite{Smeets2016} and ensure tensile intercellular stresses during tissue spreading \cite{Zimmermann2016} (\cref{Fig particles-c}). Finally, jamming due to increasing cell density and friction has also been studied using particle models \cite{Henkes2011,Garcia2015}. As in the SPV model, jammed packings of self-propelled particles exhibit collective oscillations \cite{Henkes2011} (\cref{Fig particles-d}).

\subsubsection{Discussion} \label{discussion-particle}

On the one hand, particle models miss details related to cell shape and its coupling to polarity, which are relevant for some aspects of epithelial dynamics. On the other hand, the particle description is suited to study collective migration not only of epithelial but also of mesenchymal cells, which is coordinated by weak and transient cell-cell contacts \cite{Theveneau2013}. Moreover, the description of cell-cell interactions in terms of a potential allows to compute the tissue stress tensor, thus enabling to study how and which interactions determine the tensile mechanical state of a tissue \cite{Zimmermann2016,Smeets2016,George2017}. Finally, including active cell-cell forces in the force balance \cref{eq force-balance-particles} is possible and it may lead to important insights in the future.

\subsection{Continuum models} \label{continuum}

Continuum models do not describe individual cells but set the coarse-graining level at the multicellular scale \cite{Banerjee2018}. In this approach, the cell colony is described by fields such as velocity $\vec{v}(\vec{r},t)$, polarity $\vec{p}(\vec{r},t)$, and cell density $\rho(\vec{r},t)$ that locally average these variables over many cells. Then, generic dynamical equations for the fields can be written based on the principles of hydrodynamic descriptions, observing symmetries and conservation laws. Here, we review how the dynamics of compressible polar media may be applied to model collective cell migration in epithelial monolayers. Because they contain many terms, the general equations are usually simplified to include only a few effects that are deemed most important for a particular phenomenon. We discuss the most common simplifications.


\subsubsection{Free energy of compressible polar media} \label{free-energy-continuum}

The starting point is the free energy of quiescent compressible polar media \cite{Kung2006,Voituriez2006,Cates2018}, which reads
\begin{equation} \label{eq free-energy-compressible-polar}
\mathcal{F} = \int_{\mathcal{A}} \left[\frac{\kappa}{2} (\delta\rho)^2 + \frac{a}{2} |\vec{p}|^2 + \frac{b}{4} |\vec{p}|^4 + w\, \delta\rho\,\vec{\nabla}\cdot\vec{p} + \frac{K}{2} \vec{\nabla}\vec{p}:\vec{\nabla}\vec{p} + \frac{D}{2} |\vec{\nabla}\rho|^2\right] \dd^2\vec{r}.
\end{equation}
The first term penalizes density variations $\delta\rho(\vec{r},t) = \rho (\vec{r},t) - \rho_0$ around $\rho_0$ with a bulk modulus $\kappa$ (\cref{compressibility}). The second and third terms correspond to a Landau expansion on the polarity field. The non-polarized and polarized state are stable for $a>0$ and $a<0$, respectively, with $b>0$ for stability. The fourth term couples the density and the polarity fields. In equilibrium, this term contributes a polarity $\vec{p}\propto w \vec{\nabla}\rho$, pointing towards increasing or decreasing density for $w>0$ and $w<0$, respectively. Thus, with $w<0$, this term may model interactions like CIL \cite{Marcq2014} (\cref{CRL}), which favor cell motility away from dense regions. Finally, the last two terms penalize spatial variations of the density and polarity fields, thus endowing them with a finite correlation length. In fact, the fifth term corresponds to the orientational Frank elasticity of liquid crystals in the so-called one constant approximation, which assumes that bend and splay deformations have a common modulus $K$ \cite{deGennes-Prost}. This term captures polarity alignment interactions between cells (\cref{CC-alignment}).

\subsubsection{Density and polarity dynamics} \label{density-polarity-dynamics-continuum}

Then, one writes down dynamical equations. First, cell number balance is imposed by means of a continuity equation for the density field:
\begin{equation} \label{eq density-dynamics-continuum}
\partial_t \rho + \vec{\nabla}\cdot (\rho \vec{v}) = k(\rho) \rho,
\end{equation}
where $k(\rho)$ is the net cell proliferation rate combining cell divisions and deaths \cite{Recho2016}. Second, the long-wavelength dynamics of the polarity field is given by the theory of polar media,
\begin{equation} \label{eq polarity-dynamics-continuum}
\frac{D \vec{p}}{Dt} = \frac{1}{\gamma} \vec{h} - \frac{\bar{\nu}}{2} (\vec{\nabla}\cdot\vec{v})\, \vec{p} - \nu\,\mathbf{\tilde{v}}\cdot \vec{p} + \frac{\nu_s}{\gamma}\vec{v},
\end{equation}
where we have neglected higher-order active terms \cite{Marchetti2013,Prost2015,Julicher2018}. Here, the corotational derivative of a vector $\vec{A}$ reads $D\vec{A}/Dt = (\partial_t + \vec{v}\cdot\vec{\nabla}) \vec{A} +\mathbf{\omega}\cdot\vec{A}$, where $\mathbf{\omega} = (\vec{\nabla}\vec{v} - (\vec{\nabla}\vec{v})^T)/2$ is the vorticity tensor. This derivative accounts for the advective and co-rotational transport of the polarity field. In the first term on the right-hand side of \cref{eq polarity-dynamics-continuum}, the so-called molecular field $\vec{h}= - \delta \mathcal{F}/\delta \vec{p}$ is the generalized force (torque) acting on the polarity field to minimize the free energy $\mathcal{F}$. The ensuing polarity changes are damped by the rotational friction $\gamma$, which may capture dissipation due to both cell-substrate friction (\cref{CS-friction}) and cytoskeleton reorganizations. The following two terms express the couplings of the polarity to bulk and shear flows, with coefficients $\bar{\nu}$ and $\nu$, respectively. Bulk flows are described by the velocity divergence $\vec{\nabla}\cdot\vec{v}$, whereas shear flows are described by the symmetric and traceless part of the strain rate tensor, $\mathbf{\tilde{v}} = (\vec{\nabla}\vec{v} + (\vec{\nabla}\vec{v})^T - \vec{\nabla}\cdot\vec{v}\,\mathbb{I})/2$. These terms might capture the tendency of the polarity to align with normal stresses \cite{Blanch-Mercader2017c} (plithotaxis, \cref{polarity-flow}). Finally, the last term couples the polarity to uniform flows, with coefficient $\nu_s$. This coupling might capture polarity-velocity alignment interactions (\cref{polarity-velocity}), but it has not been considered yet in continuum models of collective cell migration.

\subsubsection{Force balance} \label{force-balance-continuum}

In addition to the dynamical equations \cref{eq density-dynamics-continuum,eq polarity-dynamics-continuum}, force balance is established between the internal forces in the tissue, given in terms of its stress tensor $\mathbf{\sigma}(\vec{r},t)$, and cell-substrate (traction) forces $\vec{T}(\vec{r},t)$,
\begin{equation}
\vec{\nabla}\cdot\mathbf{\sigma} = \vec{T};\qquad \mathbf{\sigma} = - P\,\mathbb{I} + \mathbf{\sigma}^{\textrm{s}} + \mathbf{\sigma}^{\textrm{a}}.
\end{equation}
It is convenient to separate the stress tensor into the pressure $P$ and the deviatoric stress with symmetric and antisymmetric parts $\mathbf{\sigma}^{\textrm{s}}$ and $\mathbf{\sigma}^{\textrm{a}} = 1/2\,(\vec{p}\,\vec{h} - \vec{h}\,\vec{p})$. The pressure can be computed via the Gibbs-Duhem thermodynamic relation: $P = \mu\rho - f$, where $\mu=\delta \mathcal{F}/\delta\rho$ is the chemical potential and $f$ is the free energy density, namely the integrand of \cref{eq free-energy-compressible-polar}. Then, the key modeling step is to specify constitutive equations that relate the deviatoric stress tensor $\mathbf{\sigma}^{\textrm{s}}$ and the traction forces $\vec{T}$ to the velocity, polarity, and density fields, thus phenomenologically capturing cellular interactions at a coarse-grained level. Given the tissue rheology, the theory of active polar media provides generic constitutive equations
\cite{Marchetti2013,Prost2015,Julicher2018}, which we review in the following two subsections.

Different models in the literature have described migrating tissues as either elastic or fluid media, and both descriptions have successfully reproduced experimental observations. Elastic models have been recently reviewed \cite{Banerjee2018}. Here, we present the basis and general formulation of viscoelastic fluid models. Nevertheless, most of the formalism can be readily adapted to elastic descriptions by taking the limit of a long viscoelastic relaxation time $\tau\rightarrow \infty$. We discuss key differences between how elastic and fluid models describe tissue spreading in \cref{collective-phenomena-continuum}.

\subsubsection{Constitutive equation for the deviatoric stress} \label{constitutive-deviatoric-continuum}

Cell aggregates largely devoid of extracellular matrix behave as viscoelastic fluids, exhibiting an elastic response at high frequencies and a viscous response at low frequencies \cite{Gonzalez-Rodriguez2012,Khalilgharibi2016}. Thus, the simplest rheological choice is the Maxwell model, for which stress relaxes with a time-scale $\tau$.
In this case, the constitutive equation reads
\begin{multline} \label{eq stress-continuum}
\left(1 + \tau \frac{D}{Dt}\right) \left[\mathbf{\sigma}^{\textrm{s}} - \frac{\nu}{2} (\vec{p}\,\vec{h} + \vec{h}\,\vec{p} - (\vec{p}\cdot\vec{h})\,\mathbb{I}) - \bar{\nu}\, (\vec{p}\cdot\vec{h})\, \mathbb{I} + \zeta \,\mathbf{Q} + (\bar{\zeta} + \zeta' |\vec{p}|^2)\,\mathbb{I} \right] =
\\= 2\eta\, \tilde{\mathbf{v}} + \bar{\eta} \,\vec{\nabla}\cdot\vec{v}.
\end{multline}
The corotational derivative of a second rank tensor $\mathbf{A}$ reads $D\mathbf{A}/Dt = (\partial_t + \vec{v}\cdot\vec{\nabla}) \mathbf{A} + \mathbf{\omega} \cdot \mathbf{A} - \mathbf{A} \cdot \mathbf{\omega}$, where $\mathbf{\omega} = (\vec{\nabla}\vec{v} - (\vec{\nabla}\vec{v})^T)/2$ is the vorticity tensor.
The relaxation time $\tau$ is set by the processes that dominate energy dissipation. These processes may be intracellular, such as cytoskeleton reorganizations, or intercellular, such as cell-cell sliding. They are thought to release stress at time-scales of protein turnover in the cytoskeleton and in cell-cell junctions, which are of the order of tens of minutes at most \cite{Wyatt2016,Khalilgharibi2016}. In addition, other processes such as cell division, death, and extrusion \cite{Ranft2010,Matoz-Fernandez2017}, as well as cell shape fluctuations \cite{Marmottant2009,Etournay2015,Tlili2018a} and topological rearrangements \cite{Etournay2015,Krajnc2018} also fluidize the tissue over different time-scales. In general, cell migration is a really slow process, imposing strain rates of $\sim h^{-1}$, which are slower than the fastest relaxation times. Hence, migrating cell monolayers are generally expected to behave as liquids and, indeed, they exhibit liquid-like phenomena like wetting transitions \cite{Douezan2011,Gonzalez-Rodriguez2012,Perez-Gonzalez2019} and fingering instabilities \cite{Poujade2007,Alert2019} (further discussion in \cref{collective-phenomena-continuum}).

Besides the rheological model, the constitutive equation includes different types of stresses. Shear and bulk viscous stresses, proportional to the respective viscosities $\eta$ and $\bar{\eta}$, account for cell-cell friction (\cref{CC-friction}) and for the dissipation associated to density changes (\cref{compressibility}). For elastic deformations, $\eta/\tau$ and $\bar{\eta}/\tau$ are, respectively, the shear (\cref{CC-adhesion}) and bulk (\cref{compressibility}) moduli of the cell monolayer. In turn, anisotropic active stresses are proportional to the nematic tensor $\mathbf{Q} = \vec{p}\,\vec{p} - 1/2\, |\vec{p}\,|^2\,\mathbb{I}$, with coefficient $\zeta$, whereas isotropic active stresses have the coefficients $\bar{\zeta},\zeta'$ (\cref{CC-active}).
Finally, the terms with $\nu$ and $\bar{\nu}$ are the stresses associated to the flow-polarity coupling discussed above (\cref{density-polarity-dynamics-continuum,polarity-flow}).

\subsubsection{Constitutive equation for the traction forces} \label{constitutive-traction-continuum}

The next step is to specify the constitutive equation for the traction forces. Its expression is less conventional than that for the internal stress, but it was recently shown to read \cite{Oriola2017}
\begin{equation} \label{eq traction-continuum}
\vec{T} = \xi \vec{v} - T_a \vec{p} + \nu_s \dot{\vec{p}}
\end{equation}
in the long-time limit. Here, the first term is a cell-substrate viscous friction (\cref{CS-friction}), the second term is the active polar force that drives cell migration (\cref{CS-active,Fig continuum-a}), and the third term accounts for the cell-substrate forces associated to polarity-velocity alignment (\cref{polarity-velocity,eq polarity-dynamics-continuum}).

\subsubsection{Boundary conditions} \label{boundary-conditions-continuum}

In addition to the equations, boundary conditions must be specified. First, many models impose vanishing density and stress at the tissue edge. Alternatively, to capture interfacial effects, some models include a line tension \cite{Ravasio2015,Williamson2018,Alert2019}. Second, the tendency of several cell types to polarize towards free space (\cref{CRL}) can be captured by imposing perpendicular (homeotropic) anchoring of the polarity at the tissue edge \cite{Perez-Gonzalez2019,Alert2019}. In this case, for an unpolarized tissue ($a>0$ in \cref{eq free-energy-compressible-polar}), the polarity decays from the boundary with the characteristic length $L_c=\sqrt{K/a}$, which defines the width of the polarized boundary layer observed in experiments \cite{Blanch-Mercader2017,Perez-Gonzalez2019} (\cref{Fig continuum-a}). If cells align rather tangential the boundary, planar or tilted anchoring may be imposed \cite{Duclos2018}

\subsubsection{Common simplifications} \label{simplifications-continuum}

Different models have simplified the equations in different ways. Most models that focus on tissue flow do not describe the density field altogether.
Moreover, many of these models neglect flow-polarity interactions \cite{Lee2011a,Blanch-Mercader2017,Perez-Gonzalez2019,Alert2019} and even consider the polarity dynamics to be quasi-static ($\partial_t\vec{p}\approx 0$) \cite{Blanch-Mercader2017,Perez-Gonzalez2019,Alert2019}. Other models work in the limit of small correlation length of the polarity field, $L_c = \sqrt{K/a}\rightarrow 0$, whereby the polarity field drops from the description \cite{Cochet-Escartin2014,Recho2016,Yabunaka2017a,Williamson2018}. In this case, active tractions are strictly localized at the tissue edge, amounting to a non-zero boundary stress.

Another common simplification is to focus on a single dissipation source, thus only keeping either internal viscosity \cite{Douezan2012,Beaune2014,Perez-Gonzalez2019} or cell-substrate friction \cite{Cochet-Escartin2014,Tlili2018}. Friction dominates over hydrodynamic interactions above the screening length $\lambda = \sqrt{\eta/\xi}$. However, some studies suggest that this length may sometimes be comparable to tissue size, and hence both friction and internal viscosity must be kept \cite{Duclos2018,Alert2018c,Alert2019}. On a related note, some continuum models approximate two-dimensional tissue flows as incompressible \cite{Cochet-Escartin2014}. Even though it may be a valid approximation in some situations, monolayer area is not conserved (\cref{compressibility}), and hence 2D incompressibility is not a general feature of epithelial monolayers. Finally, different models have included only isotropic \cite{Banerjee2015,Notbohm2016} or anisotropic \cite{Blanch-Mercader2017c,Saw2017,Duclos2018} active stresses, or a combination of them \cite{Lee2011a,Perez-Gonzalez2019}.


\subsubsection{Collective phenomena} \label{collective-phenomena-continuum}

\begin{figure*}[tb]
\begin{center}
\includegraphics[width=\textwidth]{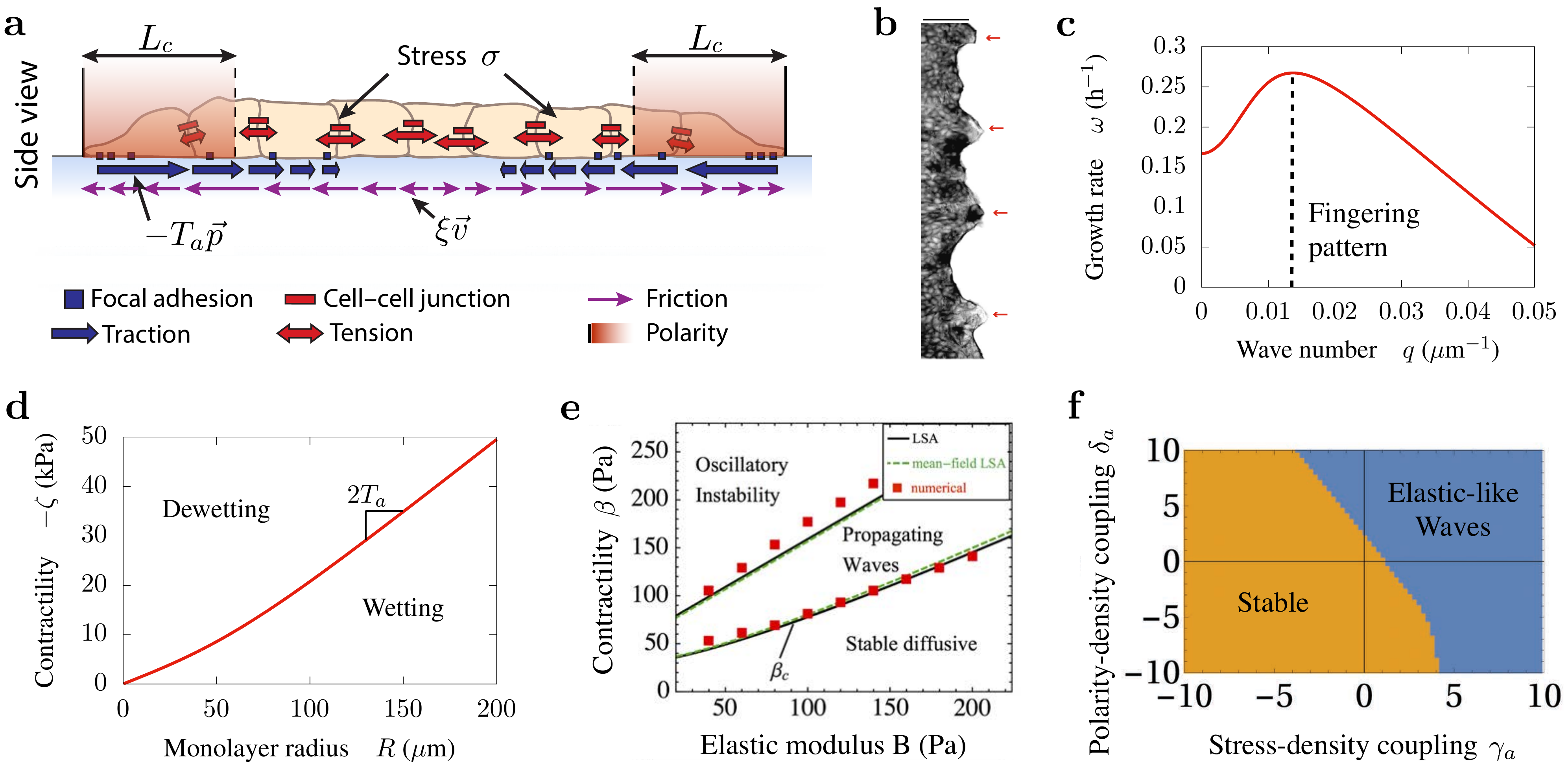}
\end{center}
  {\phantomsubcaption\label{Fig continuum-a}}
  {\phantomsubcaption\label{Fig continuum-b}}
  {\phantomsubcaption\label{Fig continuum-c}}
  {\phantomsubcaption\label{Fig continuum-d}}
  {\phantomsubcaption\label{Fig continuum-e}}
  {\phantomsubcaption\label{Fig continuum-f}}
\bfcaption{Continuum Models}{ \subref*{Fig continuum-a}, Scheme of a continuum model of a spreading monolayer. The red shade indicates the polarized boundary layer of width $L_c$. The stress $\sigma$ accounts for intercellular and intracellular tension. Adapted from \cite{Perez-Gonzalez2019}. \subref*{Fig continuum-b}, The monolayer edge develops multicellular fingers (red arrows) with a typical spacing. Scale bar, $100$ $\mu$m. Reprinted from \cite{Vishwakarma2018}. \subref*{Fig continuum-c}, Positive growth rates of tissue shape perturbations of wavenumber $q$ indicate a fingering instability. The maximal growth rate identifies the wavelength of the fingering pattern (dashed line). Adapted from \cite{Alert2019}. Copyright (2019) by the American Physical Society. \subref*{Fig continuum-d}, Phase diagram of tissue spreading. For a given contractility, monolayers larger than a critical radius spread (wetting) whereas smaller monolayers retract (dewetting). Adapted from \cite{Perez-Gonzalez2019}. \subref*{Fig continuum-e}, Stability diagram of mechanochemical waves in an elastic model. The active stress is $-\bar{\zeta}=\beta \ln(c/c_0)$, where $c$ is the concentration of a strain-regulated protein. Adapted from \cite{Banerjee2015}. Copyright (2015) by the American Physical Society. \subref*{Fig continuum-f}, Stability diagram of mechanical waves in a fluid model. $\delta_a$ adds an active contribution to the polarity-density coupling $w$ in \cref{eq free-energy-compressible-polar}, and $\gamma_a$ adds a density-dependent term to the contractility: $-\bar{\zeta} + \gamma_a \rho$. Adapted from \cite{Yabunaka2017d} by permission of The Royal Society of Chemistry.
} \label{Fig continuum}
\end{figure*}

Continuum models have focused on the spreading of epithelial monolayers, in particular addressing the formation of multicellular fingers \cite{Lee2011a,Kopf2013,BenAmar2016,Alert2019}. Consistent with experiments \cite{Vishwakarma2018} (\cref{Fig continuum-b}), a recent model has shown that, even in the absence of motility regulation at the monolayer edge, there is an active instability that leads to a fingering pattern with an intrinsic wavelength \cite{Alert2019} (\cref{Fig continuum-c}). Using the same model, recent work demonstrated a wetting transition between tissue spreading and retraction as a result of the competition between active tractions and contractile stresses. The balance between these active forces depends on monolayer size and, hence, only monolayers larger than a critical size can spread \cite{Perez-Gonzalez2019} (\cref{Fig continuum-d}). This model has been extended to account for the role of substrate stiffness on tissue spreading, hence making predictions about tissue durotaxis \cite{Alert2018c}.

A long-standing discussion in the field has been whether elastic or fluid models are more appropriate to describe tissue spreading \cite{Banerjee2018}. Whereas spreading monolayers are expected and observed to behave like liquids (see \cref{constitutive-deviatoric-continuum}), they also exhibit effective elastic responses even at long times \cite{Serra-Picamal2012,Vincent2015}, probably due to mechanotransduction processes. To capture this behavior, some models directly assume an elastic rheology of the monolayer \cite{Kopf2013,Kopf2015,Banerjee2015,Notbohm2016}, whereas others explain it based on a viscous model with time-dependent parameters \cite{Blanch-Mercader2017}. This controversy has also led to different explanations of the elastic-like mechanical waves observed during spreading \cite{Serra-Picamal2012}. Elastic models posit mechanochemical feedbacks, whereby either active stresses are coupled to the concentration of a strain-regulated protein \cite{Kopf2013,Kopf2015,Banerjee2015,Notbohm2016} (\cref{Fig continuum-e}), or the polarity is coupled to strain \cite{Tlili2018}. In contrast, fluid models rely on the combination of active forces and either flow-polarity \cite{Blanch-Mercader2017c} or density-polarity \cite{Yabunaka2017d} couplings to obtain effectively elastic waves (\cref{Fig continuum-f}).


\subsubsection{Discussion} \label{discussion-continuum}

A different type of continuum models are not based on liquid crystal physics but on the Toner-Tu equations for flocking \cite{Zimmermann2014a,Nesbitt2017,Bogdan2018}. Even though they also describe compressible polar fluids, the Toner-Tu equations do not include hydrodynamic interactions (``dry fluids'' \cite{Marchetti2013}) and do not distinguish between the polarity and the velocity fields. The facts that traction forces are observed even in static cell monolayers \cite{Mertz2012,Notbohm2016,Perez-Gonzalez2019}, and that traction and velocity fields are sometimes misaligned (kenotaxis) \cite{Kim2013,Notbohm2016}, calls for a separation of polarity and velocity. Hence, even though they may correctly capture some phenomena, flocking-type continuum models do not seem generically appropriate to describe collective cell migration.

As network models, continuum models mostly restrict to describing the migration of cohesive cell groups. The main strength of this approach is the analytical tractability of the field equations, which often allows getting analytical predictions, hence yielding insights without having to explore parameter space in simulations. However, the high degree of coarse-graining is double-edged. On the one hand, the generic equations are versatile, physically well-grounded, and can be written even without knowledge of microscopic details. On the other hand, this means that cell-cell interactions are not implemented at the cellular level but rather encoded in phenomenological couplings whose relationship to cellular processes may be unclear. Finally, at the vicinity of the jamming transition, cell shape fluctuations become a slow field that must be incorporated into hydrodynamic descriptions. To this end, Czajkowski et al. have recently proposed a model that couples a cell shape anisotropy field to the polarity field \cite{Czajkowski2018}.

\section{Conclusions and outlook} \label{conclusions}

Concurrent advances in experiments and theory are quickly shaping a solid understanding of collective cell migration. Following the footsteps of more mature areas of physics, the field can now aim towards making quantitative predictions and testing specific models and assumptions in theory-inspired experiments. Besides progress in this direction, fundamental challenges remain.

For example, here we have restricted our attention to 2D migrating cell sheets. However, as mentioned in the introduction, cells also migrate within 3D environments, which are remodeled by and mediate mechanical interactions between cells, and even provide single cells with migration modes that are unavailable in 2D\cite{Friedl2011,Bergert2015}. Whether and how cells integrate these modes and interactions to move collectively is unknown. Moreover, techniques to probe mechanical forces in 3D remain limited in accuracy and throughput\cite{Roca-Cusachs2017}. Generalizing 2D theories and techniques to 3D environments is thus one of the current challenges of cell migration biophysics.

Another central challenge is to bridge descriptions at different scales. For example, deriving continuum models from cell-scale models would map cellular interactions to tissue-scale mechanical properties and phenomena. Another pending task is to link the molecular mechanisms of actin polymerization and cell-substrate adhesion with the actual self-propulsion and friction forces included in models of collective cell migration. In particular, the most appropriate type of cell-substrate friction is still unclear \cite{Christensen2018}, largely due to the lack of experimental evidence.

A third limitation is the lack of a unifying picture that captures the mechanical consequences of cell-cell contact interactions. Upon collision, cells can adhere, repel, or ignore each other. Within each of these three behaviors there exist many nuances as well (see \cref{forces}). A better understanding of the mechanisms that define the mechanics and migration of two cells upon contact is crucial to advance our understanding of collective cell migration. A closer collaboration between physicists and life scientists is needed to incorporate the broad diversity of biological mechanisms involved in collective cell migration into a unifying physical picture.

\section*{Acknowledgments}

We apologize to the many colleagues whose work could not be cited owing to space constraints. We thank Jaume Casademunt, Carles Blanch-Mercader, and Carlos P\'{e}rez-Gonz\'{a}lez for a stimulating collaboration and many discussions. We thank Anna Labernadie for assistance with \cref{Fig interactions}, and Amin Doostmohammadi, Jos\'{e} Mu\~{n}oz, Vito Conte, and Matej Krajnc for a critical reading of the manuscript. R.A. acknowledges support from the Human Frontiers of Science Program (LT-000475/2018-C). X.T. is supported by MINECO/FEDER (BFU2015-65074-P), the Generalitat de Catalunya (2014-SGR-927 and the CERCA Programme), the European Research Council (CoG-616480), the European Commission (Grant Agreeement SEP-210342844), and Obra Social ``La Caixa''. IBEC is recipient of a Severo Ochoa Award of Excellence from the MINECO.


\bibliography{All}
\end{document}